\documentclass[runningheads]{llncs}
\usepackage[T1]{fontenc}
\usepackage{xcolor}
\PassOptionsToPackage{table}{xcolor}
\usepackage{colortbl}
\usepackage{graphicx}
\usepackage{multirow}
\usepackage{booktabs}
\usepackage{tabularx}
\usepackage{amsmath}
\usepackage{amsfonts}
\usepackage{amssymb}
\usepackage{pifont}
\usepackage{verbatim}
\usepackage{makecell}
\usepackage{hyperref}

\definecolor{mygray}{gray}{.9}

\usepackage{tikz}  

\newcommand{\imp}[1]{$_{{\textbf{\textcolor{Better}{#1}}}}$}

\definecolor{Better}{rgb}{0.18, 0.607, 0.266}

\definecolor{Worse}{rgb}{0.55, 0.0, 0.0}

\newcommand{\xx}{\mathbf{x}}

\newcommand{\yy}{\mathbf{y}}

\newcommand{\mypar}[1]{\vspace{3pt}\noindent\textbf{#1}}

\newcommand{\Ours}{\textsc{Reflect}}

\usepackage{graphicx,verbatim}

\begin{document}

\title{\Ours{}: \underline{Re}ctified \underline{Fl}ows for \underline{E}fficient Brain Anomaly \underline{C}orrection \underline{T}ransport.}

\author{Farzad Beizaee\inst{1,2} \and
Sina Hajimiri\inst{1} \and
Ismail Ben Ayed\inst{1} \and \\
Gregory Lodygensky\inst{2} \and
Christian Desrosiers\inst{1} \and
Jose Dolz\inst{1}}


%
\authorrunning{F. Beizaee et al.}
%
\institute{ÉTS Montreal, Canada \and
CHU Sainte-Justine, University of Montreal, Canada\\
\email{farzad.beizaee.1@ens.etsmtl.ca}}

\titlerunning{Rectified Flows for Efficient Brain
Anomaly Correction Transport.}
\maketitle              
\begin{abstract}
Unsupervised anomaly detection (UAD) in brain imaging is crucial for identifying pathologies without the need for labeled data. However, accurately localizing anomalies remains challenging due to the intricate structure of brain anatomy and the scarcity of abnormal examples. In this work, we introduce \Ours{}, a novel framework that leverages rectified flows to establish a direct, linear trajectory for correcting abnormal MR images toward a normal distribution. By learning a straight, one-step correction transport map, our method efficiently corrects brain anomalies and can precisely localize anomalies by detecting discrepancies between anomalous input and corrected counterpart. In contrast to the diffusion-based UAD models, which require iterative stochastic sampling, rectified flows provide a direct transport map, enabling single-step inference. Extensive experiments on popular UAD brain segmentation benchmarks demonstrate that \Ours{} significantly outperforms state-of-the-art unsupervised anomaly detection methods. The code is available at \href{https://github.com/farzad-bz/REFLECT}{https://github.com/farzad-bz/REFLECT}.


\end{abstract}

\section{Introduction}

Brain anomaly detection using medical images is a critical task in neuroimaging, with significant implications for early diagnosis and treatment planning. Brain abnormalities such as tumors, lesions, or traumatic injuries often appear as structural deviations from normal anatomy. While supervised methods for localizing anomalies are effective in many settings, they rely on large annotated datasets, which are costly and often scarce, particularly for rare anomalies. This has led to growing interest in unsupervised methods, where models learn the normal data distribution and subsequently localize deviations as potential abnormalities. 

In recent years, generative models have been widely investigated for \linebreak reconstruction-based unsupervised anomaly detection (UAD). Initial attempts leverage Auto-Encoders (AEs)~\cite{baur2021autoencoders,baur2019deep} and their variants, including Variational Auto-Encoders (VAEs) \cite{silva2022constrained,zimmerer2019unsupervised}. Generative Adversarial Networks (GANs) \cite{goodfellow2014gan}, including AnoGAN~\cite{schlegl2017unsupervised} and f-AnoGAN~\cite{schlegl2019fast}, have also emerged as promising alternatives to AE-based methods. However, these approaches tend to overfit the normal training data or yield blurry reconstructions. Normalizing flows (NFs)~\cite{tabak2013family,rezende2015variational} are another family of generative models that transform a simple base distribution into a complex target distribution through a series of invertible transformations, thereby allowing exact likelihood estimation. This makes them particularly appealing for anomaly localization via out-of-distribution detection~\cite {gudovskiy2022cflow,kim2023sanflow,chiu2023self,zhao2023ae}.
However, NFs require complex architectures, are computationally expensive, and often involve iterative steps.

Progress in generative modeling has led to the rise of diffusion models \cite{ho2020denoising}, a powerful class of probabilistic models that generate high-quality data by gradually transforming noise into structured outputs through a learned iterative process. Fueled by their impressive generative performance, diffusion models have been increasingly adopted in medical imaging tasks, including unsupervised anomaly detection \cite{behrendt2023patched,behrendt2024leveraging,beizaee2025mad,bercea2024diffusion,liang2024itermask,marimont2024disyre,naval2024ensembled,wyatt2022anoddpm}. AnoDDPM~\cite{wyatt2022anoddpm} resorts to a partial diffusion strategy, where it adds noise to the image up to a certain timestep and then recovers it via reverse diffusion, whereas pDDPM~\cite{behrendt2023patched} applies diffusion patch-wise to better capture local context.  
THOR~\cite{bercea2024diffusion} refines diffusion models by using implicit temporal guidance via anomaly maps during the reconstruction process. Inspired by diffusion models, Itermask~\cite{liang2024itermask} proposes iterative mask refinement using reconstruction errors to better localize brain lesions in MRI. And very recently, MAD-AD~\cite{beizaee2025mad} treats abnormalities as noise in the latent space and uses a masked diffusion process to selectively correct abnormalities.  While diffusion models offer a better overall performance, they tend to ``memorize'' patterns from training data~\cite{somepalli2023understanding}, which reduces their generalizability. Also, they need many iterative steps to reconstruct normal images, even with DDIM sampling~\cite{songdenoising}. Last but not least, all these generative methods are primarily designed to generate new samples (often from pure noise), and not to modify existing images. This limits their applicability for the selective correction of images. 

An alternative unsupervised approach uses self-supervised learning to restore normal images corrupted with synthetic anomalies. Methods like Foreign Patch Interpolation (FPI)\cite{tan2022detecting} and Poisson Image Interpolation (PII)\cite{tan2021detecting} introduce such defects by blending patches from different normal images, enabling pixel-level anomaly localization. Most recently, DISYRE v2~\cite{naval2024ensembled} introduced a cold-diffusion pipeline that restores synthetically corrupted images with controlled anomaly severity through iterative refinement. While these methods are effective, their reliance on synthetic anomalies limits generalization, as these may not fully capture the variability of real-world abnormalities.

To address the aforementioned limitations, we propose \Ours{}, an unsupervised brain anomaly detection framework built on the recently introduced rectified flows \cite{liu2023flow}. Rectified flows, which learn a transport map between two different distributions through rectified trajectories, exhibit several advantages over the previously mentioned generative models, such as improved stability, high-fidelity reconstructions, and direct and efficient mapping, which enables the mapping of a sample to a target distribution with a single step. 

Our work makes the following key contributions: \emph{i}) We propose leveraging rectified flows in latent space to enable optimal and straight transport of abnormal brain samples toward their normal counterparts, thereby facilitating unsupervised brain anomaly detection with enhanced accuracy and reliability. To the best of our knowledge, rectified flow has not been explored before for unsupervised anomaly detection. By learning straight flow trajectories, our approach requires fewer time steps, which enables high-quality correction of abnormal regions in a single step while preserving normal regions. 
Unlike the often complex formulations used in diffusion models or GANs, rectified flows' objective directly controls the geometry of the flow and provides a natural mechanism to correct anomalous samples, making the method both theoretically elegant and practically efficient. \emph{ii}) Moreover, we introduce an effective technique for generating diverse and realistic anomalous brain MRIs using a random walk-based masking strategy operating in the latent space. Masked regions are replaced with textured image segments or random noise, enhancing the variability and realism of synthesized anomalies. \emph{iii}) Extensive experiments on brain anomaly detection benchmarks further demonstrate the superiority of our approach over recent state-of-the-art unsupervised anomaly detection methods.

\section{Preliminaries: Rectified Flows}
\label{sec:prelim}

Rectified flows~\cite{liu2023flow}, building upon the principles of continuous normalizing flows \cite{chen2018neural}, learn an ordinary differential equation (ODE) that transports samples from an initial distribution \(\pi_0\) to a target distribution \(\pi_1\) along trajectories that are as straight as possible in a continuous-time framework. Concretely, consider the linear interpolation between random variables \(X_0 \sim \pi_0\) and \(X_1 \sim \pi_1\):
\begin{equation}
    X_t = (1-t)\,X_0 + t\,X_1, \quad t \in [0,1].
    \label{eq:linear-interpolation}
\end{equation}
Although \(X_t\) provides a continuous trajectory from \(X_0\) to \(X_1\), these straight-line paths are non-causal and may intersect when considering different sample pairs. Such intersections are undesirable for generative modeling, as they lead to non-causal and non-deterministic dynamics. This interpolation is “rectified” into a causal ODE flow $\{ Z_t: t \in [0,1] \}$ via
\begin{equation}
    \label{eq:rectified_flow}
    \frac{dZ_t}{dt} \;=\; v(Z_t, t), \quad Z_0 \sim \pi_0,
\end{equation}
where $Z_{t}$ represents the state of the rectified flow at time $t$, and $v(\cdot,t)$ is a trainable velocity field. We seek $v$ that best aligns with the direction of linear interpolation paths.
Therefore, the training objective minimizes the discrepancy between the true instantaneous velocity of the interpolation and the learned velocity:
\begin{equation}
    \min_{v} \; \int_{0}^{1} \mathbb{E}\Big[\| (X_1 - X_0) - v_\theta(X_t, t) \|^2\Big]\,dt.
    \label{eq:rf-loss}
\end{equation}
In practice, we approximate the integral via Monte Carlo sampling:
\begin{equation}
    \min_{\theta} \;\; \mathbb{E}_{t \sim \mathrm{U}[0,1]} 
    \Big[\big\| (X_1 - X_0) - v_\theta \big(X_t,\; t\big)\big\|^2\Big].
\end{equation}

Moreover, the \emph{reflow} procedure, an iterative application of rectified flow training to the outputs of a previous flow, can further straighten trajectories and minimize discretization errors. This process helps prevent trajectory crossings induced by paired input samples and further refines and rectifies the transport paths toward the target distribution, thereby enhancing robustness and enabling high-quality single-step sampling while preserving distribution fidelity. This is achieved through the following objective:
\begin{equation}
    \min_{\theta} \;\; \mathbb{E}_{Z_0 \sim \pi_0, t \sim \mathrm{U}(0,1)} \Big[\big\|(Z_1 - Z_0)-v'_\theta(Z_t, t) \big\|^2\Big] , \quad Z_0 = X_0,
\end{equation}
where $Z_1$ represents the transported sample ($Z_0$) with the initial rectified flow.

\begin{figure}[t!]
    \centering
    \includegraphics[width=\linewidth]{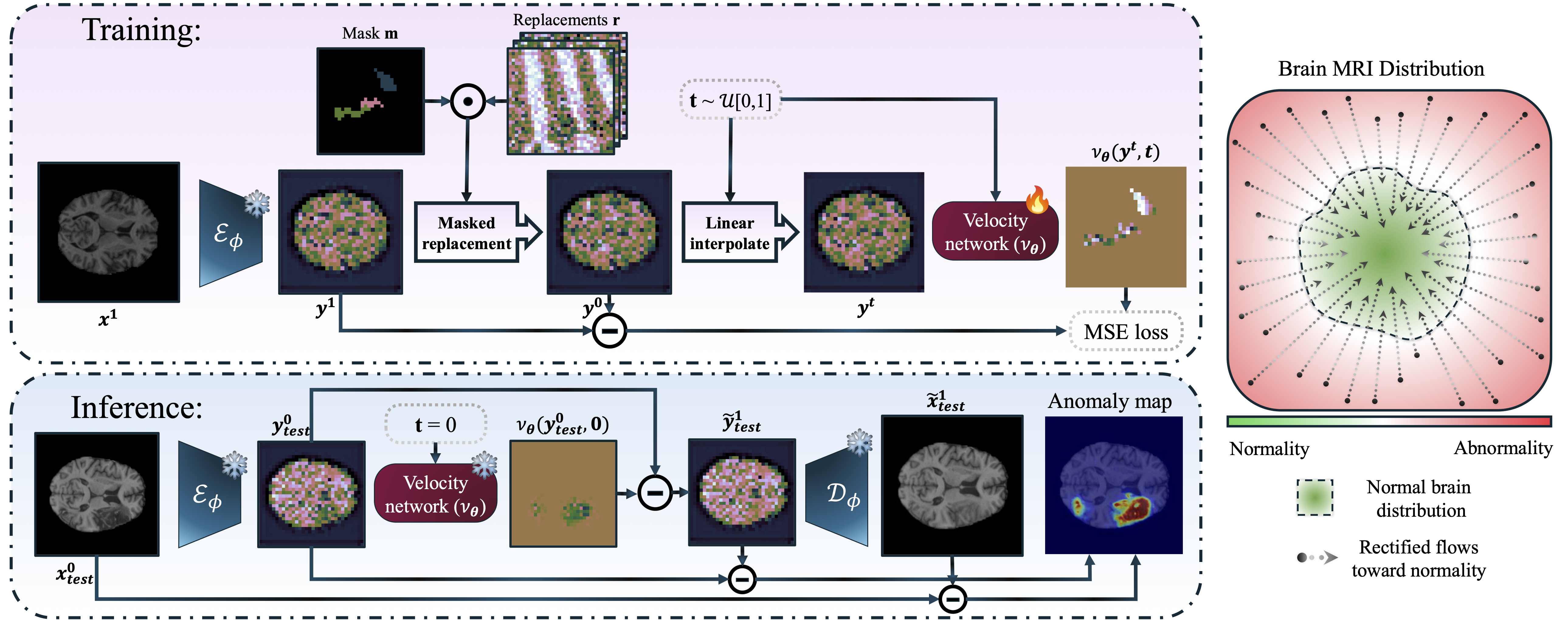}
    \caption{\textbf{Overview of \Ours{}}. \textit{Training:} a velocity network is trained to predict the displacement between the latent representations of normal images and their %
    corrupted versions, 
    using their linear interpolation and time $t$. \textit{Inference:} Given an anomalous test image, the velocity network computes the flow, allowing the image to be corrected in a single step. \textit{Right:} Rectified flows toward normality in the brain's latent space. }
    \label{fig:method-overview}
\end{figure}

\section{Method}
Interpreting $\pi_0$ as the entire brain distribution encompassing both normal and abnormal (or injured) brains, and $\pi_1$ as the distribution of normal brains (with $\pi_1 \subset \pi_0$), we train a rectified flow to learn a velocity field $v_\theta(., t)$ that transports abnormal samples from $\pi_0$ toward 
$\pi_1$ (illustrated in Fig. \ref{fig:method-overview}), while leaving normal samples unchanged since they already belong to $\pi_1$.
Our framework for unsupervised brain anomaly detection using rectified flows (see Figure \ref{fig:method-overview}) consists of three main stages: (1) generating paired abnormal and normal images in the latent space, (2) rectified flow training to transport abnormal brains toward normality, and (3) localizing anomalies. Below, we describe each stage in detail.

\subsection{Generating Paired Samples}
Following \cite{esser2024scaling,beizaee2025mad}, we transform the data into a latent space to stabilize training, improve output quality, and enable the interpretation of anomalies as noise or out-of-distribution features. To achieve this, we adapt and fine-tune a pre-trained VAE~\cite{rombach2022high} for medical images, which compresses high-dimensional data into a compact latent representation while retaining structures and semantic information. Let $\mathcal{X}=\{\mathbf{x}^{(i)}\}_{i=1}^N$ denote the set of normal brain samples; the encoder $\mathcal{E}_\phi$ maps each image to a latent code via $\mathbf{y}^{(i)} = \mathcal{E}_\phi(\mathbf{x}^{(i)})$.

To train the rectified flow for transforming anomalous latent representations to normal ones, we construct paired latent vectors $(\mathbf{y}_i^0, \mathbf{y}_i^1)$, where $\mathbf{y}_i^1$ is a normal sample and $\mathbf{y}_i^0$ is an artificially corrupted version. The corruption is applied using a binary mask $\mathbf{m}$ that specifies the regions to be altered, a random replacement vector $\mathbf{r}$ indicating the direction of corruption, and parameter $\alpha \in [0,1]$ which controls the severity of corruption. The corrupted latent vector is given by:
\begin{equation}
\label{eq:corruption}
\yy^0_i \, = \, \left(\sqrt{1 - \alpha} \cdot \yy^1_i \,+\, \sqrt{\alpha} \cdot \mathbf{r}\right) \odot \mathbf{m} \,+\, \yy^1_i \odot (1 - \mathbf{m}),
\end{equation}
which guarantees that if both $\mathbf{y}_i^1$ and $\mathbf{r}$ follow a standard normal distribution, the resulting corrupted image will also follow a standard normal distribution.

The introduced corruptions must closely mimic genuine injuries and span the entire range of possible anomalies. We assume that the brain may exhibit up to $N$ distinct anomalous regions delineated by non-overlapping masks. The mask of each region is generated by a random walk starting from a random point within the brain and taking a random number of uniformly probable steps to neighboring points. Masked positions then correspond to the points visited during the walk. This strategy results in more realistic anomaly shapes than rectangular patch-based masking \cite{beizaee2025mad}. 

Afterward, a replacement vector is assigned to each masked region, and the final corrupted image is generated according to Eq.~\ref{eq:corruption}. For each region, the replacement vector is generated by randomly choosing between two distinct strategies. The first strategy is to use random noise, motivated by the interpretation of anomalies as noise in the latent space. Alternatively, we can use a cropped segment from the latent representation of a textured image (not necessarily medical images) to impose more realistic and structured anomalies. For the first strategy, in our implementation, we propose sampling random noise independently for each channel according to:
\begin{equation}
\mathbf{r} \, = \, \sqrt{\beta} \cdot q \, +\, \sqrt{1 - \beta} \cdot \mathbf{p},
\end{equation}
where $q \in \mathbb{R} \sim \mathcal{N}(0, 1)$, $\mathbf{p} \in \mathbb{R}^{H \times W} \sim \mathcal{N}(\mathbf{0}, \mathbf{I})$, and $\beta$ is uniformly sampled from $[0, 1]$ (fixed for all channels). This formulation can be interpreted as a weighted combination of a global (image-level) random variable and a spatially varying (pixel-wise) random vector, constructed such that the resulting $\mathbf{r}$ follows a standard normal distribution. This strategy imposes spatial dependencies and yields more realistic and structured corruptions in the latent space.

\subsection{Training Rectified Flows for Anomaly Correction}

Our goal is to learn a mapping that transports abnormal latent samples $\yy^0 \sim \pi_0$ back to the normal distribution $\pi_1$. For this purpose, we train the velocity field $v_{\theta}(.,t)$ using the paired latent data $(\yy^0, \yy^1)$. The training objective becomes:
\begin{equation}
    \min_{\theta} \;\; \mathbb{E}_{t \sim \mathrm{U}[0,1]} 
    \Big[\big\| (Y_1 - Y_0) - v_\theta \big(Y_t,\; t\big)\big\|^2\Big],
\end{equation}
where $Y_t = (1-t)\, Y_0 + t\, Y_1 $. Once the velocity model is trained, the learned ODE (Eq. \ref{eq:rectified_flow}) defines a flow that can be numerically simulated using a standard ODE solver (e.g., Euler method). Note that in many cases, including ours, a coarse discretization (even a single Euler step) is sufficient due to the straightening effect of the learned flow. Furthermore, after obtaining the first velocity model, we can refine the flow by training a second rectified flow model $v'(.,\, t)$ using the reflow process. This additional stage could further straighten the flows and enhance the correction trajectories toward the target normal distribution (i.e., anomaly-free brains). We refer to this model as 2-\Ours{}, while the first model is denoted as 1-\Ours{}.

\subsection{Inference and Anomaly Localization}
At test time, given an image $\xx^0_\text{test}$ (potentially containing anomalies), we first encode it as $\yy^0_\text{test} = \mathcal{E_{\phi}}(\xx^0_\text{test})$. 
Then, we solve the reverse-time ODE with the trained velocity model using a single Euler step, which results in:
\begin{equation}
    \tilde{\yy}^1_\text{test} \, = \, \yy^0_\text{test} -\,v_\theta(\yy^0_\text{test},\, 0),
\end{equation}
where $\tilde{\yy}^1_\text{test}$ should ideally lie within the $\pi_1$ distribution. The reconstructed normal image is decoded from the corrected latent space using the VAE decoder: $\tilde{\xx}^1_\text{test} = \mathcal{D}_{\phi}(\tilde{\yy}^1_\text{test})$. Finally, anomalies are localized by comparing the original image to its reconstruction in both latent space and image space:
\begin{equation}
    \mathbf{A}(\mathbf{x}_\text{test}) \, = \, \frac{1}{2} \left|\tilde{\xx}^1_\text{test} - \xx^0_\text{test}\right| \, + \, \frac{1}{2} \left|\tilde{\yy}^1_\text{test} - \yy^0_\text{test}\right|.
\end{equation}
Higher differences indicate regions where the flow had to make 
larger corrections, thereby signaling anomalies.

\section{Experiments}

\subsection{Experimental Setting}

\mypar{Datasets.} For our experiments, we employed BraTS'21~\cite{bakas2017advancing} that comprises 1251 brain scans across four modalities (T1, Contrast-Enhanced T1 (T1CE), T2, and FLAIR), and ATLAS 2.0~\cite{liew2022large}, which contains 655 T1-weighted MRI scans. Both datasets come with expert-annotated lesion masks. We extracted 20 central axial slices from skull-stripped brains and padded them to a resolution of 256×256 pixels. Both datasets are divided into training (80\%), validation (10\%), and testing (10\%) subsets, and only normal training slices are used for training, while the single slice of the test subjects displaying the most prominent pathology is reserved for inference. Moreover, the Describable Textures Dataset (DTD)~\cite{cimpoi2014describing} is used and converted to gray-scale to serve as textured replacement images.

\mypar{Evaluation metrics.} 
Following \cite{beizaee2025mad}, we assess models' performance using the Maximum Dice score, which reflects the highest Dice coefficient achieved as the threshold varies from 0 to 1. 

\mypar{Implementation details.} 
In each iteration, the number of masked regions is randomly chosen between 1 and 4, and the number of random-walk steps is randomly sampled between 0 and 200. \mbox{1-\Ours{}} model underwent training for 200 epochs for BraTS'21 and for 400 epochs for the ATLAS dataset with a batch size of 96, using the AdamW optimizer and a learning rate of $5\times 10^{-4}$. Afterward, 2-\Ours{} was trained on top of 1-\Ours{} for another 50 epochs with a learning rate of $1\times 10^{-5}$. Also, we have used 5 reverse ODE correction steps.

\begin{table}[t!]
\setlength{\tabcolsep}{6pt}
\centering
    \caption{\textbf{Quantitative results} obtained by different approaches. The best method per modality and/or dataset is highlighted in \textbf{bold}, whereas the second one is \underline{underlined}. The performance gains over the best baseline are shown in \textcolor{Better}{green}.}
\label{table:quantitative}
\resizebox{\textwidth}{!}{
\begin{tabular}{l@{\hskip 2px}l|c|cccc}
\toprule 
\multicolumn{1}{l}{\multirow[b]{2}{*}{Method}} & & \multicolumn{1}{c|}{ATLAS 2.0}  & \multicolumn{4}{c}{BraTS'21} \\
 & & T1-w & FLAIR & T1CE & T2-w & T1-w\\
\midrule
AE~\cite{baur2021autoencoders} & $_{\textsl {MedIA'21}}$ & 11.9 & 33.4 & 32.3 & 30.2 & 28.5  \\
DDPM~\cite{ho2020denoising} & $_{\textsl {Neurips'20}}$ & 20.2 & 60.7 & 37.9& 36.4  & 29.4 \\
AutoDDPM~\cite{bercea2023mask} & $_{\textsl {ICML Workshop'23}}$ & 12.7 & 55.5 & 36.9 & 29.7 & 33.5  \\
DAE~\cite{kascenas2022denoising} & $_{\textsl {MIDL'22}}$ & 11.1 & 79.7  & 36.7  & 69.6 & 29.5  \\
Cycl.UNet~\cite{liang2023modality} & $_{\textsl {MICCAI'23}}$ & N/A & 65.0 & 42.6 & 49.5 & 37.0  \\
IterMask$^2$~\cite{liang2024itermask} & $_{\textsl {MICCAI'24}}$ & 35.3 & \underline{80.2} & 61.7 & 71.2 & 58.5 \\

MAD-AD~\cite{beizaee2025mad} & $_{\textsl {IPMI'25}}$  & \underline{36.1} & 76.2 & \underline{68.5} & \underline{73.2} & \underline{63.4} \\
\rowcolor{mygray}
\textbf{1-\Ours} & $_{\textsl {Ours}}$
& \textbf{41.6}\imp{+5.5} 
& \textbf{85.1}\imp{+4.9} 
& \textbf{73.0}\imp{+4.5} 
& 79.6\imp{+6.4} 
&  69.7\imp{+6.3} 
\\
\rowcolor{mygray}
\textbf{2-\Ours} & $_{\textsl {Ours}}$
& 40.8 \imp{+4.7} 
& 83.2\imp{+3.0} 
& 72.0\imp{+3.5} 
&  \textbf{80.3}\imp{+7.1} 
& \textbf{69.8}\imp{+6.4}\\
\bottomrule
\end{tabular}
}
\end{table}

\subsection{Results}
\mypar{Main quantitative results.} We empirically evaluate the performance of the proposed approach compared to relevant UAD methods proposed for brain MRI, whose results are reported in Table \ref{table:quantitative}. These values demonstrate that \Ours{} substantially outperforms recent state-of-the-art brain UAD methods. Compared to the second-best approach, i.e., MAD-AD~\cite{beizaee2025mad}, our method brings improvement gains ranging from 4.5\% to 6.4\% in both datasets. In particular, in T1-w, which seems to be the most difficult modality based on baselines' results, \Ours{} yields the highest difference gaps (similar to gains in T2-w). These results demonstrate the superiority of our method across modalities and datasets, compared to very relevant baselines. 
Regarding \textbf{reflow}'s effect (i.e., 2-\Ours{}) on the first trained model 1-\Ours{}, results (Table \ref{table:quantitative}) suggest that the reflow process did not improve the overall performance of the anomaly detection, likely because the initial flows are already well-rectified, and the reflow process is attached to the anomaly localization performance of the first method.

\begin{figure}[t!]
    \centering
    \includegraphics[width=\linewidth]{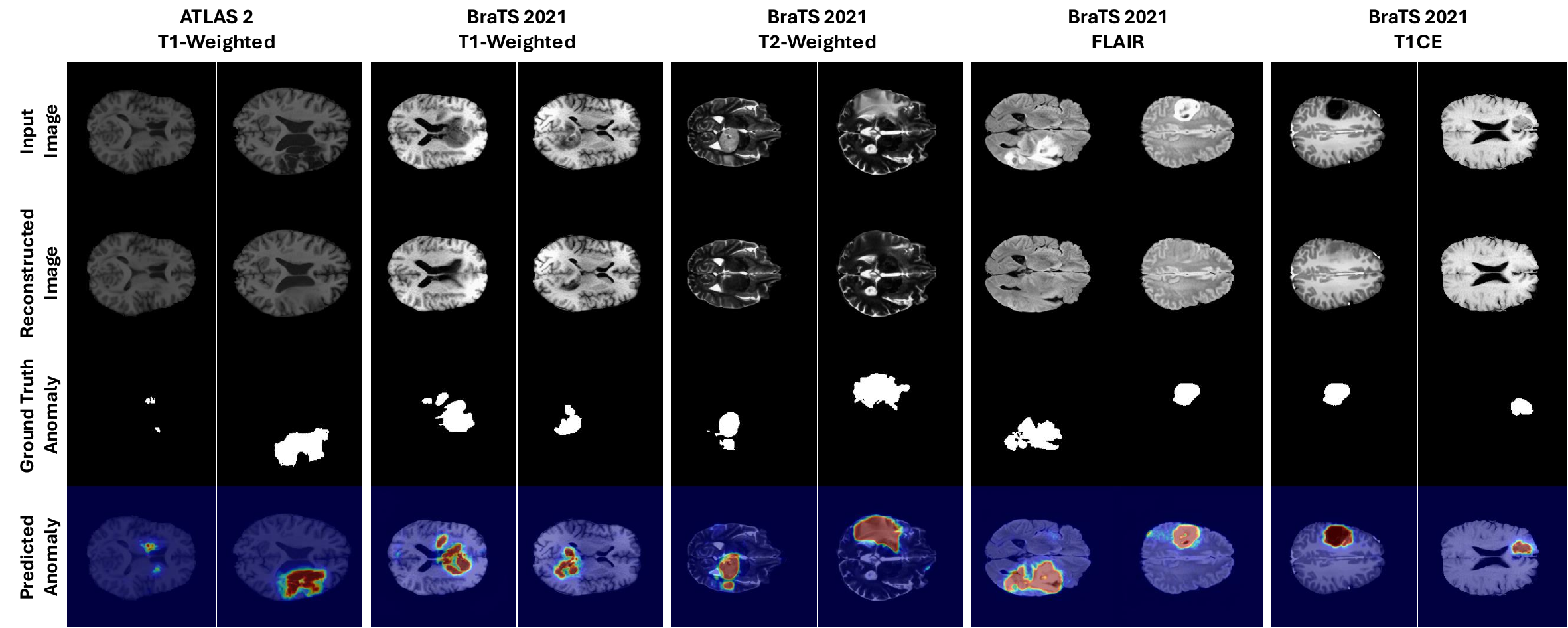}
    \caption{\textbf{Qualitative results.} \textit{Second row:} Reconstructed images of their abnormal input counterparts. \textit{Last row:} Anomaly segmentation maps obtained by our approach.} 
    \label{fig:qualitative}
\end{figure}

\begin{figure}[t!]
    \centering
    \includegraphics[width=\linewidth]{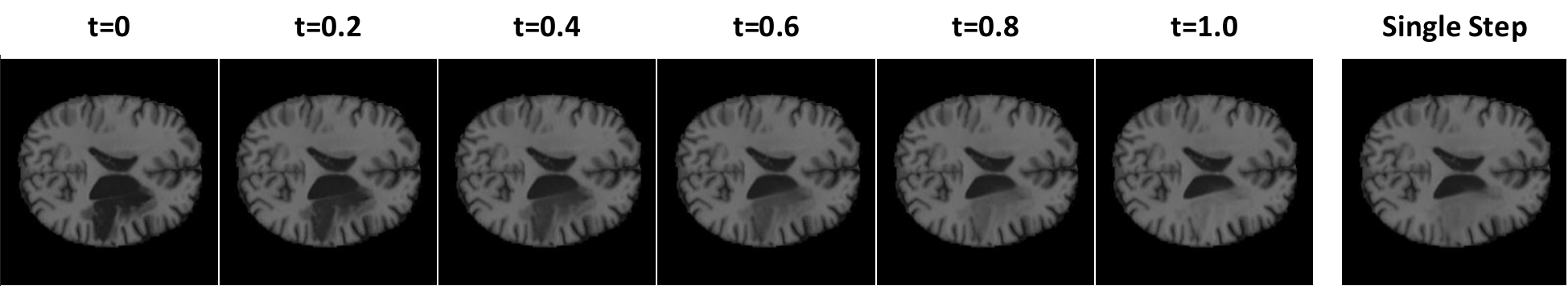}
    \caption{Transition between an anomalous to healthy brain using 10 reverse ODE steps \textit{vs.} a single step, showing that flows are well-rectified and effective in a single step.}
    \label{fig:transport-visualization}
\end{figure}

\mypar{Qualitative results.} To visually assess the effectiveness of \Ours{}, we depict in Fig. \ref{fig:qualitative} several visual examples of both the reconstructed images (\textit{second row}) and the predicted anomaly maps (\textit{last row}) across all modalities of the BraTS dataset. As these images highlight, \Ours{} successfully reconstructs realistic ``healthy'' images, leading to accurately predicted anomaly maps. Fig.~\ref{fig:transport-visualization} depicts a visual example of the transitions performed by the rectified flow.

\mypar{Ablations.} \textit{Model size:} We evaluated five different model sizes to analyze the trade-off between model efficiency and anomaly detection performance.
While larger models demonstrated slightly better performance, smaller variants still achieved highly competitive results, significantly outperforming previous methods, making \Ours{}  a strong choice for real-time or resource-constrained applications. \textit{Effect of the VAE model:} We also investigated the effect of the VAE model by comparing two variants with scale factors of 4 and 8. As illustrated in Table~\ref{tab:ablation-studies}, both VAEs perform well, with the VAE using a scale factor of 4 exhibiting slightly better performance, likely due to its higher embedding dimensionality. However, this comes at the cost of increased computational requirements.

\begin{table}[t!]
    \centering
\setlength{\tabcolsep}{6pt}
\scriptsize
\caption{Ablation studies (BRATS'21) on model size and VAE model employed.}
\label{tab:ablation-studies}
\begin{tabular}{ll|c|cccc|c}
\toprule
& \multirow[b]{2}{*}{Model} & \multirow[b]{2}{*}{\#Params} &  \multicolumn{4}{c}{Modality}\\
\cmidrule(l{5pt}r{5pt}){4-8}
 & & & FLAIR & T1CE & T2-w & T1-w & Avg \\
 \midrule
\multirow{5}{*}{\rotatebox[origin=c]{90}{Model size}} & 
UNet XS & $\sim$16 M & 81.7 & 70.6 & 80.9 & 69.7 & 75.7 \\
 & UNet S & $\sim$64 M & 83.9 & 72.1 & 81.0 & 69.6 & 76.7 \\
 & UNet M* & $\sim$145 M & \textbf{85.1} & 73.0 & 79.6 & 69.7 & 76.9 \\
 & UNet L & $\sim$257 M & 83.2 & 72.1 & \textbf{80.8} & \textbf{71.2} & 76.8\\
 & UNet XL & $\sim$580 M & 83.6 & \textbf{73.3} & 79.9 & 70.7 & \textbf{76.9} \\
\midrule
 \multirow{2}{*}{\rotatebox[origin=c]{90}{VAE}} 
 & KL-f4 & $\sim$55 M & \textbf{85.4} & 72.1 & \textbf{80.8} & \textbf{71.8} & \textbf{77.5} \\
 & KL-f8* & $\sim$84 M & 85.1 & \textbf{73.0} & 79.6 & 69.7 & 76.9 \\
\bottomrule
\end{tabular}
\end{table}

\section{Conclusion}
We introduced \Ours{}, an unsupervised framework that leverages rectified flows in latent space, which provides a direct one-step correction transport that significantly improves reconstruction fidelity and anomaly localization. Experimental results on established unsupervised anomaly detection benchmarks confirm that \Ours{} outperforms current state-of-the-art approaches, paving the way for more robust and efficient diagnostic tools in neuroimaging. Future directions include extending the framework to 3D data and testing cross-dataset generalization to boost clinical relevance.

\subsubsection{\discintname}
The authors have no competing interests to declare that are relevant to the content of this article.

\bibliographystyle{splncs04}
\bibliography{mybibliography}

\end{document}